\documentclass[final,3p,12pt,times,onecolumn]{elsarticle}

\usepackage{graphicx}
\usepackage{epsfig}
\usepackage{amssymb}
\usepackage{mathrsfs}
\usepackage{bm}
\usepackage{color}
\usepackage{amsmath}
\usepackage{mathrsfs}
\usepackage{algorithm}
\usepackage{algorithmic}
\usepackage{mathtools}
\biboptions{numbers,sort&compress}

\usepackage{geometry}
\usepackage{graphicx}
\usepackage{subfigure}
\usepackage{float}

\usepackage{threeparttable}

\usepackage{epstopdf}
\usepackage{amsmath}
\usepackage{blindtext}

\usepackage{amsmath}
\usepackage{bbm}
\usepackage{float}
\usepackage{subfigure}
\usepackage{url}

\usepackage{multirow}

\journal{Cell Reports Physical Science}

\begin{document}
\begin{frontmatter}

\title{Device-independent Verification of Quantum Coherence without Quantum Control}

\author{Yan-Han Yang$^1$
\corref{equal}}
\author{Xue Yang$^{2,1}$
\corref{equal}}
\author{Xing-Zhou Zheng$^1$}

\author{Ming-Xing Luo$^{1,3,4}$ \corref{correspondingauthor}
}
\address{$^{1}$ School of Information Science and Technology, Southwest Jiaotong University, Chengdu 610031, China}
\address{$^{2}$ School of Computer and Network Security, Chengdu University of Technology, Chengdu 610059, China}
\address{$^{3}$ CAS Center for Excellence in Quantum Information and Quantum Physics, Hefei, 230026, China}
\address{$^{4}$ Lead contact: mxluo@swjtu.edu.cn}

\cortext[equal]
{These authors contributed equally}
\cortext[correspondingauthor]{Corresponding author: mxluo@swjtu.edu.cn}

\end{frontmatter}

\section*{\Large Summary}

Quantum coherence plays a crucial role in manipulating and controlling quantum systems, leading to breakthroughs in various fields such as quantum information, quantum sensing, and the detection of gravitational waves. Most coherence witnesses rely on the assumption of being able to control quantum states. Here we report a device-independent coherence model by extending the standard Bell theory to multiple source scenarios. We propose a Greenberger-Horne-Zeilinger-type paradox to verify the particle and wave behaviors of a coherent carrier. We experimentally generate generalized two-photon entangled states that violate the present paradox, witnessing spatial quantum superposition through local measurements.

\section*{\Large Introduction}

The quantum superposition principle allows linearly combining two or more quantum states to create a new and valid quantum state. This underlies various distinguishing features of quantum mechanics, including quantum entanglement$^{1-4}$ and quantum tunneling$^{5,6}$. The so-called quantum coherence sets quantum systems apart from classical systems and has been recognized as highly significant in the advancement of quantum information and technologies$^7$. The degree of coherence in such systems can be quantified to assess their potential for applications in various fields, including cryptography$^{8-11}$, quantum computation$^{12,13}$, metrology$^{14}$, detection of gravitational waves$^{15}$, and thermodynamics$^{16,17}$.

State tomography has traditionally been employed as a direct method$^{18,19}$ to detect coherence in experimental setups. This requires several measurements and is then time-consuming and resource-intensive. Another operational approach makes use of coherence witnesses$^{20-24}$ by ruling out all incoherent states. Coherence witness provides an interesting way to assess the degree of coherence in a system without requiring a complete characterization of the quantum state, making it applicable in experimental studies$^{25-28}$. Both methods depend on the assumptions of the trusted quantum devices and then not device-independent.

Bell inequality provides a device-independent method to verify coherence features. Bell theory was originally proposed to witness the nonlocal correlations arise from quantum systems, which does not depend on specific quantum devices but only on the measurement statistics$^{1,29,30}$. In a standard Bell test experiment, one source distributes the state to two separated observers who perform local measurements. The statistical correlations depend on the shared source that supposes the independent and identically distributed (IID) assumption$^{31,32}$. The coherence is related to two distinguished physical features of wave and particle in one quantum source$^{33,34}$. Instead of a one-source experiment, each quantum feature can be verified by using different quantum sources. This further motivates Wheeler delay-choice experiment with photons$^{35-41}$ by violating specific Bell inequality. Others are based on superconducting quantum circuit$^{42,43}$, single-photon$^{44,45}$ or atoms$^{46-48}$. All of these results make use of single source assumption but emphasize the controlling feature of quantum devices in the casual model$^{36-38,49}$.

This paper proposes a way to verify quantum coherence in a device-independent (DI) model. Taking into account both particle and wave properties, we focus on the role of the source in satisfying the multiple-independent and identically distributed (IID) experimental trials, drawing inspiration from Young double-slit experiment and Wheeler delay-choice experiment. We propose a generalized  Greenberger-Horne-Zeilinger (GHZ) paradox$^{50}$ captures the correlations that arise from quantum coherent systems ruling out all classical incoherent sources. We further extend this concept to communication tasks and coherence equality within causal models in a prepare-and-measure scenario$^{49,51}$. We finally propose a two-photon experiment utilizing spatial quantum superposition as a means to verify coherence.

\begin{figure}[!ht]
\begin{center}
\includegraphics[width=0.5\linewidth]{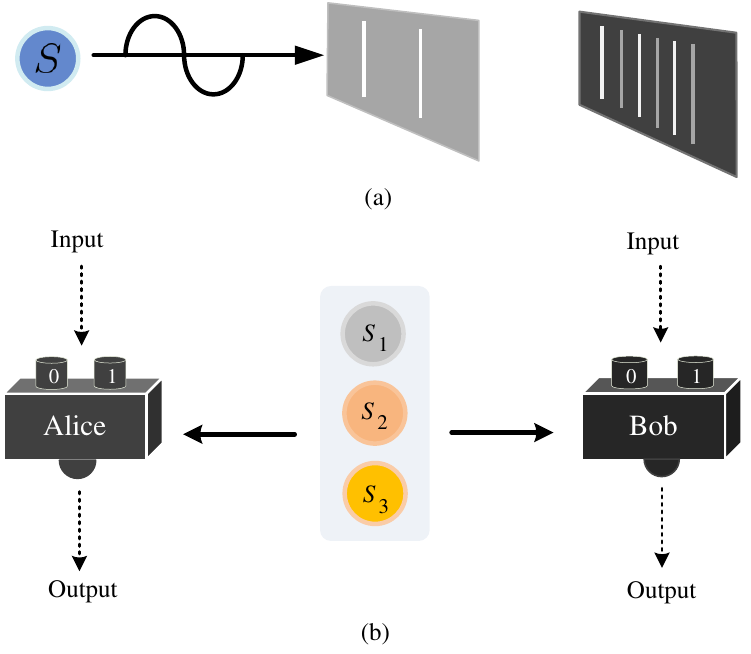}
\caption{Young two-slit experiment and our proposed device-independent coherence model. (a) A beam of particles emitted by one source $S$ is directed onto a barrier with two slits and then arrives at a screen to observe the pattern formed by the particles. (b) The device-independent coherence model incorporates multiple sources (e.g., $S_1, S_2, S_3$). The measurement inputs for two parties are labeled as $x\in\{0,1\}$ and $y\in\{0,1\}$, respectively, and the corresponding outputs are labeled $a$ and $b$. }
\label{figure1}
\end{center}
\end{figure}

\section*{\Large Results}

\section*{Device-independent coherence model}

Consider a standard Bell scenario. One source distributes two physical states to distant observers. Each observer measures the received state depending on local measurement type $x$ or $y$ to output an integer $a$ or $b$. The joint distribution $P(a,b|x,y)$ of outcomes conditional on the measurement types in the local hidden variable (LHV) model allows the decomposition as$^{52}$:
\begin{equation}
P(a,b|x,y)=\int_\Omega
d\mu(\lambda) p(a|x,\lambda)p(b|y,\lambda),
\label{eqbell}
\end{equation}
where $(\Omega, \mu(\lambda))$ is the measure space of the variable $\lambda$. Violating this decomposition for experimental correlations represents the quantum nonlocality in Bell experiments.

For verifying the coherence, there are three different scenarios of states arriving at the detectors in both Young double-slit experiment$^{33}$ and Wheeler delay-choice experiment$^{34,49}$, i.e., one particle passing through either of two different paths and one wave passing through two paths simultaneously (Figure \ref{figure1}(a)). As both features should be verified in a single reasonable experiment, this intrigues a natural assumption of multiple sources beyond the standard Bell scenario in a device-independent coherence model (Figure \ref{figure1}(b)), where both detectors can be space-separated. Similar to Bell's theory the joint probability distribution in this scenario can be decomposed into
\begin{equation}
P_i(a,b|x,y)=\int d\mu(\lambda_i) p(a|x,\lambda_i)p(b|y,\lambda_i),
\label{eqn02}
\end{equation}
for each source $\lambda_i$ ($i\in\{1,\dots, m\}$). This implies a general method for verifying the quantum coherence using a set of Bell-like inequalities$^{29,52}$ as
\begin{equation}
E(P_i(a,b|x,y))\leq C_i,
\label{eqclass}
\end{equation}
where $E(\cdot)$ is a general function and $C_i$ denotes the bound for incoherent states with the source $\lambda_i$. Here, for each source, it is an IID-experimental trial. Further restrictions will be assigned to all sources for exhibiting both the particle and wave behaviors of quantum coherence. Informally, the quantum wave source is in the linear superposition of all particles, see the discussions with GHZ paradox.

\section*{Witnessing Coherence with a GHZ-type paradox}

Greenberger, Horne, and Zeilinger-(GHZ) have provided a novel experimental method in quantum mechanics that highlights the counter-intuitive nature of quantum many-body systems$^{50,53}$.  The so-called GHZ paradox verifies the existence of multiple nonlocal correlations that cannot be explained by any LHV model. Inspired by the GHZ idea, we propose a generalized paradox to witness the quantum coherence. We first introduce the idea of GHZ paradoxes with a three-qubit maximally entangled  GHZ state$^{50}$: $|\Psi\rangle= (|000\rangle+|111\rangle)/\sqrt{2}$. From the Born's rule, the GHZ state has stabilizers consisting of Pauli matrices $\sigma_X$ and $\sigma_Y$ as
\begin{eqnarray}
\left\{
\begin{split}
 &\langle  \sigma_X^1 \sigma_Y^2 \sigma_Y^3\rangle={\rm tr}(\sigma_X^1 \sigma_Y^2 \sigma_Y^3 |\Psi\rangle\langle \Psi|)= -1,
\\
 &\langle  \sigma_Y^1 \sigma_X^2 \sigma_Y^3\rangle={\rm tr}(\sigma_Y^1 \sigma_X^2 \sigma_Y^3|\Psi\rangle\langle \Psi|)=-1,
\\
 &\langle  \sigma_Y^1 \sigma_Y^2 \sigma_X^3\rangle={\rm tr}(\sigma_Y^1 \sigma_Y^2 \sigma_X^3|\Psi\rangle\langle \Psi|)=-1,
 \\
& \langle \sigma_X^1 \sigma_X^2 \sigma_X^3\rangle={\rm tr}(\sigma_X^1 \sigma_X^2\sigma_X^3|\Psi\rangle\langle \Psi|)=+1.
\end{split}
\right.
\end{eqnarray}
According to any local hidden variable (LHV) theory, the measurement outcomes of the operators are predetermined. This means there are some variables $v_{X(Y)}^i$ corresponding to four measurements above such that $v_X^1 v_Y^2 v_Y^3=-1$, $v_Y^1 v_X^2 v_Y^3=-1$,
$v_Y^1 v_Y^2 v_X^3=-1$, and $v_X^1 v_X^2 v_X^3=1$, where $v_{X(Y)}^i\in \{\pm 1\}$. Multiplying both sides of these four equations together implies a sharp contradiction of $-1=+1$, where the left-hand side is $-1$ and the right side is $+1$.

In the extension of the GHZ paradox, we start by considering an explicit example using the maximally entangled Einstein-Podolsky-Rosen (EPR) state$^{1}$: $|\psi_{00}\rangle=(|01\rangle+|10\rangle)/\sqrt{2}$. Let us focus on the scenario (Figure \ref{figure1}(b)). With this setup, the input state is given by $|\psi_{xy}\rangle$ conditional on the experimental configuration $xy$ to identify different sources. Similar to a standard Bell test of single source$^{52}$ there are two measurement setups per party. This implies a generalized GHZ-type paradox as
\begin{eqnarray}
\left\{
\begin{aligned}
&\langle \sigma_Z^1 \sigma_Z^2 \rangle_{|\psi_{01}\rangle}={\rm tr}(\sigma_Z^1 \sigma_Z^2 |\psi_{01}\rangle\langle \psi_{01}|)=-1,
\\
&\langle \sigma_Z^1 \sigma_Z^2 \rangle_{|\psi_{10}\rangle}={\rm tr}(\sigma_Z^1 \sigma_Z^2 |\psi_{10}\rangle\langle \psi_{10}|)=-1,
\\
& \langle \sigma_X^1 \sigma_X^2\rangle_{|\psi_{01}\rangle}={\rm tr}(\sigma_X^1 \sigma_X^2 |\psi_{01}\rangle\langle \psi_{01}|)=0,
\\
&\langle  \sigma_X^1 \sigma_X^2\rangle_{|\psi_{10}\rangle}={\rm tr}(\sigma_X^1\sigma_X^2 |\psi_{10}\rangle\langle \psi_{10}|)=0,
\\
&\langle  \sigma_X^1 \sigma_X^2\rangle_{|\psi_{00}\rangle}={\rm tr}(\sigma_X^1 \sigma_X^2 |\psi_{00}\rangle\langle \psi_{00}|)=+1,
\end{aligned}
\right.
\label{eqpd01}
\end{eqnarray}
which cannot be interpreted in any LHV model. Especially, assume that the source $\lambda_{00}$ is a classical mixture of two sources $\lambda_{01}$ and $\lambda_{10}$, that is, $\lambda_{00}=p_1\lambda_{01}+p_2\lambda_{10}$ with a a given probability distribution $\{p_1,p_2\}$. From Eq.~(\ref{eqn02}), both local responding functions $p(a|x,\lambda_i)$ and $p(b|y,\lambda_i)$ are equivalent to measurable functions that linearly depend on the variable $\lambda_i$. This implies the result of $v^1v^2(xy=00)=0$ from the equality of $v^1v^2(xy=01)=0$ and  $v^1v^2(xy=10)=0$. It yields a sharp contradiction of $0=+1$ according to the last equality. This means the correlations (\ref{eqpd01}) consist of a generalized GHZ-type paradox, differing from the previous paradoxes that arise from the conflict of $-1\neq +1$, that is, the deterministic correlation from the definite value$^{54-57}$.

The present paradox (\ref{eqpd01}) can be used to verify the quantum coherence. Especially, the correlations of $\langle \sigma_Z^1 \sigma_Z^2 \rangle=-1$ and $\langle \sigma_X^1 \sigma_X^2 \rangle=0$ with respect to the states $|\psi_{01}\rangle$ and $|\psi_{01}\rangle$ allow to witness the particle behavior while the correlation of $\langle \sigma_X^1 \sigma_X^2 \rangle=+1$ with respect to
the state $|\psi_{00}\rangle$ identifies the wave behavior. This means the present paradox provides an efficient way to verify quantum coherence with both particle and wave behaviors in a device-independent manner. We further extend the GHZ paradox with other measurements to witness the coherence in multiple-slit interference experiments in SI$^{58}$.

\section*{Coherence equality with a quantum XOR game}

Quantum XOR game provides one way to quantify the discrepancy between classical and quantum scenarios$^{63}$. One special example is the coherence game$^{51}$ which can be used to identify quantum resources beyond classical variables. It depends on a coherence term defined in the LHV model by$^{51}$:
\begin{equation}
I_{ab}^{C}\equiv \sum_{x,y=0}^1 (-1)^{x\oplus y} P(a,b|x,y)=0,
\label{Iab}
\end{equation}
where $\oplus$ denotes the sum of two numbers modulo 2. Any violation of this equality witnesses an information carrier with the coherence. This can be explained as a cooperating nonlocal game, where one referee $R$ encodes random inputs $x$ and $y$ by opening or closing the slit, that is, the quantum controller of the source$^{36,39-41}$. The referee requires two players (detectors) to send two answers $a$ and $b$, respectively. Both players win the game if their outcomes satisfy the consistent condition:
\begin{equation}
a\oplus b = x\oplus y
\label{game}
\end{equation}
for any $xy\in \{00,01,10,11\}$. Combing with Eq.~(\ref{Iab}) we get the average winning probability as$^{51}$:
\begin{eqnarray}
 P_{win} &=& \frac{1}{4}[P(a=b|x=y)+P(a\not=b|x\not=y)]
\nonumber\\
 &=& \frac{1}{4}[ P(0,0|0,0)+P(0,0|1,1)+P(1,1|0,0)+P(1,1|1,1)
\nonumber\\
 && +P(0,1|0,1)+P(0,1|1,0)+P(1,0|0,1)+P(1,0|1,0) ]
\nonumber\\
 &=& \frac{1}{2}+\frac{1}{4}(I_{00}+I_{11}).
\end{eqnarray}
This implies the winning probability $P_{win}^{C}=1/2$ for the case of two players sharing a classical source from the equation (\ref{Iab}).

In the quantum scenario, suppose that there are three states as $|\psi_{xy}\rangle$ with $x\not=y$, or $x=y=0$. Denote the following states for different inputs as
\begin{eqnarray}
  |\psi_{01}\rangle &=& |0\rangle_A |1\rangle_B,
  \\
  |\psi_{10}\rangle &=& |1\rangle_A |0\rangle_B,
  \\
|\psi_{00}\rangle &=& \cos\theta|0\rangle_A |1\rangle_B +\sin\theta|1\rangle_A |0\rangle_B.
\end{eqnarray}
Let $M_{A(B)}$ be two given dichotomic observables satisfying $M_{A(B)}^2=\mathbbm{1}$ with the identity operator $\mathbbm{1}$. From the Born's rule the quantum probability of the measurement outcomes $(a,b)$ has the form of $P(a,b|x,y)={\rm tr}((M_{a|A}\otimes M_{b|B})\rho_{xy})$ for the input state $\rho_{xy}$, where $M_{a|A}$ and $M_{b|B}$ are positive semi-definite operators that satisfy $M_{a=0|A}-M_{a=1|A}=M_A$ and $M_{b=0|B}-M_{b=1|B}=M_B$. For the input $xy=00$, the average winning probability for two quantum players is given by
\begin{eqnarray}
P(a=b|0,0)&=&P(a=b=0|x=0,y=0)+P(a=b=1|x=0,y=0)
\nonumber\\
&=&
\frac{1}{2}(1+{\rm tr}((M_A\otimes M_B)|\psi_{00}\rangle\langle \psi_{00}|)).
\end{eqnarray}
For each input $xy$ with $x\not=y$, the winning probability is given by
\begin{eqnarray}
P(a\not=b|x,y)
&=&P(a=0,b=1|x,y)+P(a=1,b=0|x,y)
\nonumber\\
&=&\frac{1}{2}(1-{\rm tr}((M_A\otimes M_B)|\psi_{xy}\rangle\langle \psi_{xy}|)).
\end{eqnarray}
For the input $xy=11$, both players output random bits. This implies the total winning probability given by
\begin{eqnarray}\label{Pwin}
P_{win}&=&\frac{1}{4}(P(a=b|x=0,y=0)+P(a\not=b|x=0,y=1)
\nonumber
\\
&&
+P(a\not=b|x=1,y=0)+P(a=b|x=1,y=1))
\nonumber
\\
&=&
\frac{1}{2}+\frac{1}{8}{\rm tr}((M_A\otimes M_B)|\psi_{00}\rangle\langle \psi_{00}|) -\frac{1}{8}{\rm tr}((M_A\otimes M_B)|\psi_{01}\rangle\langle \psi_{01}|)
\nonumber
\\
&&-\frac{1}{8}{\rm tr}((M_A\otimes M_B)|\psi_{10}\rangle\langle \psi_{10}|).
\label{eqpw}
\end{eqnarray}
It follows the optimal quantum winning probability as
\begin{eqnarray}
P_{win}^{Q}=\frac{1}{2}+\frac{\sin(2\theta)}{8}
\end{eqnarray}
by choosing Pauli observable $M_A=-M_B= \sigma_X$, or $P_{win}^{Q}=5/8$ with $M_A=M_B=\sigma_Z$. According to Eq.~\eqref{Iab}, we get the maximal quantum coherent quantity as $I_{ab}^Q=(-1)^{a\oplus b}\sin(2\theta)/4$, or $I_{ab}^Q=(-1)^{a\oplus b}/4$ for this kind of local measurements.

\begin{figure}[!ht]
\centering
\includegraphics[width=0.7\linewidth]{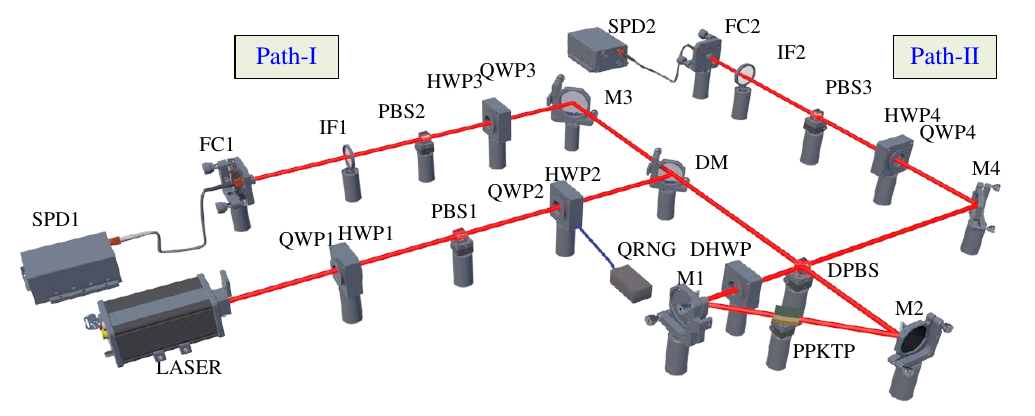}
\caption{Schematic experiment setup.  We generate orthogonally polarized photon-pairs at 810 nm wavelength with a PPKTP crystal in a Sagnac interferometer and, by erasing the ``which path''  information of the pump with PBS2. Key elements include half-wave plate (HWP), quarter-wave plate (QWP), polarizing beam splitter (PBS), quantum random-number generator (QRNG), dual-wavelength polarization beam splitter (DPBS), mirror (M), dichromatic mirror (DM), dual-wavelength half-wave plate (DHWP), interference filter (IF), fiber coupler (FC), and single photon detector (SPD). }
\label{Figure2}
\end{figure}

\section*{ Experimental result}

To verify the coherence paradox and the related coherence equality, we first prepare a set of two-photon entangled states $|\psi_{00}(\theta)\rangle=\cos\theta|\psi_{01}\rangle+\sin\theta|\psi_{10}\rangle$ with $|\psi_{01}\rangle\equiv|HV\rangle$ and $|\psi_{10}\rangle\equiv |VH\rangle$. We encode horizontal polarization state $|H\rangle$ (or vertical polarization state $|V\rangle$) of photons as qubits $|0\rangle$ (or $|1\rangle$)$^{64}$. Polarimetric entangled photons are generated using the spontaneous parametric down-conversion (SPDC) source.

The density matrices of all the prepared initial states are reconstructed by using the state tomography$^{60,65}$. The average fidelities of the six prepared states are ${F}(|\psi_{01}\rangle)=0.9973\pm0.0004$, ${F}(|\psi_{00}({\pi}/{12})\rangle)=0.9946\pm0.0008$, ${F}(|\psi_{00}({\pi}/{8})\rangle)=0.9686\pm0.0013$, ${F}(|\psi_{00}({\pi}/{6})\rangle)=0.9771\pm0.0008$, ${F}(|\psi_{00}({\pi}/{4})\rangle)=0.9937\pm0.0017$, and ${F}(|\psi_{10}\rangle)=0.9939\pm0.0001$, where the fidelity measure of the state $\rho$ with respect to the ideal state $\rho_0$ is defined by$^{2}$:
\begin{eqnarray}
F(\rho)=\rm{tr}\sqrt{\sqrt{\rho}\rho_0\sqrt{\rho}}.
\end{eqnarray}
Tomographic results of their density matrices are shown SI (Figure S1)$^{58}$. The error bars are estimated by considering the Poissonian counting statistics$^{60,62}$, indicating the uncertainty associated with the measurement.

\begin{figure}[!ht]
\centering
\includegraphics[width=0.6\linewidth]{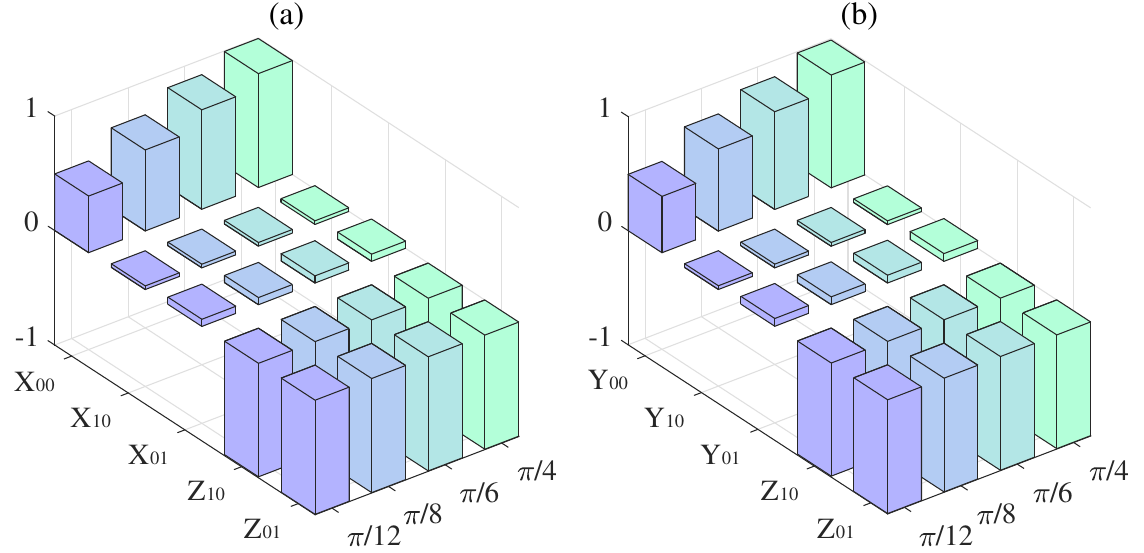}
\caption{Experimental results of the coherence paradox. (a) All the correlations used in the paradox (\ref{eqpd01}) under the measurement of $\sigma_X^1\sigma^2_X$ and $\sigma_Z^1\sigma^2_Z$. (b) All the correlations used in the paradox (\ref{eqpd01}) under the measurement of $\sigma_Y^1\sigma^2_Y$ and $\sigma_Z^1\sigma^2_Z$. There are four different states of $|\psi_{00}(\theta)\rangle$ by setting $\theta$ to $\pi/12$, $\pi/8$, $\pi/6$, and $\pi/4$, which result in four paradoxes. The labels ${\rm{X}}_{xy}$, ${\rm{Y}}_{xy}$, and ${\rm{Z}}_{xy}$ are corresponding to the experimental values of $\langle \sigma_X^1 \sigma_X^2\rangle_{|\psi_{xy}\rangle}$, $\langle \sigma_Y^1 \sigma_Y^2 \rangle_{|\psi_{xy}\rangle}$, and $\langle \sigma_Z^1 \sigma_Z^2\rangle_{|\psi_{xy}\rangle}$, respectively. }
\label{Figure3}
\end{figure}

\begin{figure}[!ht]
\begin{center}
\includegraphics[width=0.6\linewidth]{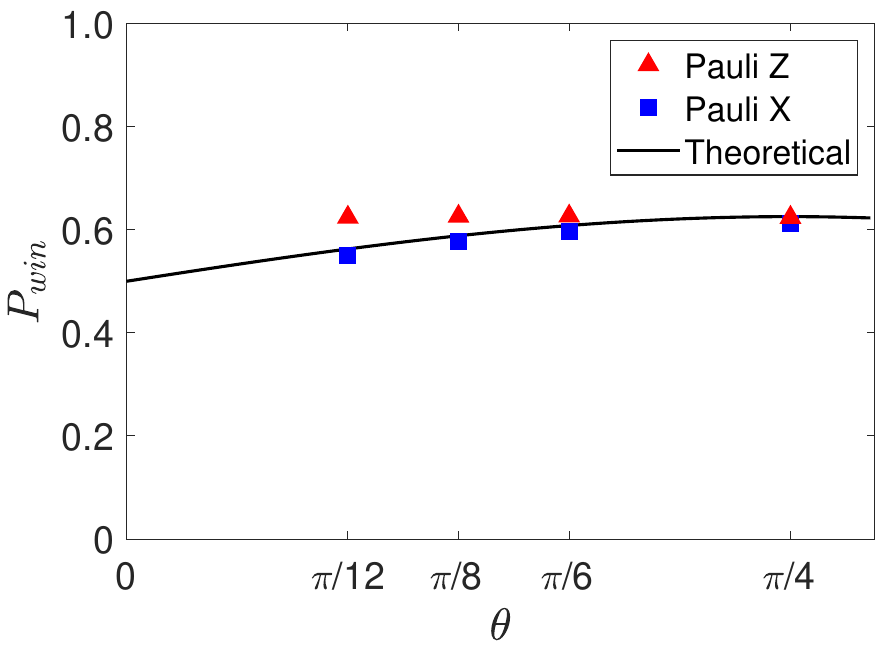}
\caption{Winning probability of the coherence game under quantum strategies. The red triangle denotes the winning probabilities that are evaluated by using the experimental outcomes based on the local measurement of Pauli $\sigma_Z$. The blue square denotes the winning probabilities that are evaluated by using the experimental outcomes based on the local measurement of Pauli $\sigma_X$. The black line denotes the theoretical winning probability. }
\label{Figure4}
\end{center}
\end{figure}

Table \ref{paradox1} shows the experimental results regarding the coherence paradox under the post-selection of four sets of photon coincidences, according to Eq.~(\ref{eqpd01}). Despite the presence of noise, the experimental values effectively confirm that the maximally entangled two-photon EPR state violates the coherence paradox. Especially, the experimental value of $\langle \sigma_X^1 \sigma_X^2\rangle_{|\psi_{00}({\pi}/{4})\rangle}$ is larger than 10 times of both correlations $\langle \sigma_X^1 \sigma_X^2\rangle_{|\psi_{01}\rangle}$ and $\langle \sigma_X^1 \sigma_X^2\rangle_{|\psi_{10}\rangle}$. This fact cannot be explained in any LHV model by following the proof procedure of the paradox (\ref{eqpd01}), that is, the classical mixture of the measurement results of independent states cannot represent the measured results of the superposition states. The $p$-value in the experiment is smaller than $10^{-15}$ which guarantees the results of experimental verification. The experimental data of generalized states are shown Table S1 and S2 in SI$^{58}$ (Figure \ref{Figure3}). All the experimental results are consistent with the theoretical predictions (S1) and (S3) in SI$^{58}$.

\begin{table}[]
    \centering
        \caption{Experimental values related to the coherence paradox of the maximally entangled two-photon EPR state.}
    \begin{tabular}{l|c|c|c}
    \hline
       Correlators  & Experimental value & Theoretical value & Validity   \\ \hline
       $\langle \sigma_Z^1 \sigma_Z^2\rangle_{|\psi_{01}\rangle}$  & $ -0.9967 \pm 0.0005$ & -1 & \checkmark
\\
        $\langle \sigma_Z^1 \sigma_Z^2\rangle_{|\psi_{10}\rangle}$  & $ -0.9912 \pm 0.0007$ & -1 & \checkmark
\\
         $\langle \sigma_X^1 \sigma_X^2\rangle_{|\psi_{01}\rangle}$  & $ 0.0625 \pm 0.0084$ & 0 & \checkmark
\\
         $\langle \sigma_X^1 \sigma_X^2\rangle_{|\psi_{10}\rangle}$  & $ 0.0317 \pm 0.0080$ & 0 & \checkmark
\\
         $\langle \sigma_X^1 \sigma_X^2\rangle_{|\psi_{00}({\pi}/{4})\rangle}$  & $0.9949 \pm 0.0006$ & 1 & \checkmark \\       \hline
    \end{tabular}
    \label{paradox1}
\end{table}

The experimental winning probability of the coherence game is evaluated by using photon correlations (\ref{correlator}) according to Eq.~(\ref{eqpw}) (Figure \ref{Figure4}). The theoretical winning probability of two quantum players can reach the maximal violation $P_{win}^{Q}=5/8$ going the classical winning probability beyond $P_{win}^{C}=1/2$ when two parties measure the received states with either Pauli matrix $\sigma_X$ or $\sigma_Z$. The red triangle represents the winning probability based on the experimental values using local measurement of Pauli $\sigma_Z$, which is independent of the angle $\theta$ and remains at the maximum violation. The blue square labels the winning probability based on the experimental values using the local measurement of  Pauli $\sigma_X$, which is consistent with the theoretical relationship. We get the maximal violation with the maximally entangled EPR state, witnessing a quantum superposition beyond any classical correlations shared by two players. The experimental data of the coherence game are shown Tables S3 and S4 in SI$^{58}$.  We declare that the present experiment does not close the locality loophole and the measurement loophole that are encountered in most Bell-type experiments$^{66}$.

\section*{\Large Discussion}

In classical physics, interference results from the superposition of waves. The phenomenon of single-particle interference closely resembles its classical counterpart. quantum coherence, which serves as evidence of the superposition principle, exhibits both wave-like and particle-like behaviors that go beyond what is observed in classical systems. In multiparticle systems, superposition leads to even more complex phenomena compared to single-particle systems. A compelling demonstration of this effect was conducted by$^{67}$, who performed elegant experiments with two-photon states. These experiments revealed an interference pattern in the correlation between the two photons, while no such pattern emerged when only one photon was observed.

Most of the existing experiments involving single-photon or two-photon entanglement rely on the assumption of quantum control$^{44-48}$, specifically controlling the photon source to induce wave-like behavior in the particles. One typical example is the delay-choice experiment with photons by controlling a single plate$^{36-41,68}$. In contrast, our coherence witness experiment relies on the generalized GHZ-type paradox (\ref{eqpd01}) which does not make any assumptions about quantum control, going beyond the state tomography$^{18,19}$ and coherence witnesses$^{20-24}$. Given that two-photon interference cannot be fully explained within the framework of single-photon systems, our experiment with coherence witnessing establishes the genuinely quantum nature of the observed optical effect concerning the generalized GHZ paradox. This discovery raises intriguing possibilities for exploring non-classical features$^{52,59}$ or communication tasks$^{51}$.

\section*{\Large Experimental procedures}

\section*{\small Resource Availability}

\subsubsection*{Lead Contact}

Further information and requests for resources should be directed to the lead contact Ming-Xing Luo (mxluo@swjtu.edu.cn).

\subsubsection*{Materials available}

This paper did not generate new materials.

\subsubsection*{Data and code availability}

The experimental data are available from the corresponding author upon reasonable request. This paper did not report original code.

\section*{Experimental setup}

In the experiment, the light source module contains a 405nm continuous-wave diode laser (Laser), a polarization beam splitter (PBS1), and two groups of wave plates (Figure \ref{Figure2}). The first wave plate group that consists of one quarter-wave plate and one half-wave plate
(QWP1+HWP1) with PBS1 is to adjust the phase and intensity of the laser, and the second wave plate group (QWP2+HWP2) regulates the intensity distribution in the Sagnac interference that creates a polarization superposition state represented by $\cos\theta |H\rangle+e^{i\varphi}\sin\theta|V\rangle$. The photons from path-I and -II are further coupled into a single-mode fiber, respectively, forming a polarization-entangled Bell-like state. The photon measurement module comprises two identical optical paths, each including an 810nm quarter-wave plate, an 810nm HWP, an 810nm PBS, and an interference filter. Our entanglement source employs beam-like type-II phase matching, which achieves high brightness (0.34MHz), high fidelity ($98\%$), and high collection efficiency ($60\%$) at the same time. The details of the experimental setup are shown in SI$^{58}$. The measured coincidence rates of visibility are illustrated in SI (Figure S2)$^{58}$.

\section*{Experimental method}

For two-photon entangled states $|\psi_{00}(\theta)\rangle=\cos\theta|\psi_{01}\rangle+\sin\theta|\psi_{10}\rangle$, the parameter $\theta\in (0,\pi/2)$ characterizes the degree of entanglement. By adjusting the parameter $\theta$ using the axis direction of HWP2 and QWP2 (Figure \ref{Figure2}), we generate a collection of photon states $\{|\psi_{01}\rangle$, $|\psi_{00}({\pi}/{12})\rangle$, $|\psi_{00}({\pi}/{8})\rangle$, $|\psi_{00}({\pi}/{6})\rangle$, $|\psi_{00}({\pi}/{4})\rangle$, $|\psi_{10}\rangle\}$.

To account for the measurement outcomes related to the coherence paradox, we begin by constructing a projection measurement basis. We represent the eigenstates of the Pauli operator $M_u$ as $|\varphi_u^a\rangle$ with $M_u\in \{\sigma_X, \sigma_Y, \sigma_Z\}$, and $u=0$ or 1. The projection basis is then expressed as $\{M_{a|u}\equiv |\varphi_u^a\rangle \langle\varphi_u^a|\}$. This is experimentally realized by adjusting the optical axis angles of four plates, HWP3+QWP3 and HWP4+QWP4 (Figure \ref{Figure2}). We perform measurements $\{M_{a|u}\}$ $(M_u\in \{\sigma_X,\sigma_Y,\sigma_Z\})$ on the photons in the path-I, and $\{M_{b|v}\}$ $(M_v\in \{\sigma_X,\sigma_Y,\sigma_Z\})$ measurements on the photons in the path-II. Denote $N_{u,v}^{a,b}$ as the coincidence photon number of the outcomes $a$ and $b$ conditional on the measurement inputs $u$ and $v$, which further represents a correlator as
\begin{eqnarray}
\langle M_u^1 M_v^2\rangle\equiv \frac{N_{u,v}^{0,0}-N_{u,v}^{0,1}-N_{u,v}^{1,0}+N_{u,v}^{1,1}}{N_{u,v}^{0,0}+N_{u,v}^{0,1}+N_{u,v}^{1,0}+N_{u,v}^{1,1}}.
\label{correlator}
\end{eqnarray}
In experiments, each measurement set took 100s to complete the evaluation. We evaluate all 18 combinations by taking into account all $10$ trials.

The standard deviation was initially utilized as a metric to quantify the statistical significance of a Bell inequality violation$^{69,70}$. Here, we estimate the $p$-value under the Poissonian distribution in statistical analysis, which does not depend on the assumption of a Gaussian distribution and the independence of each trial result for evaluating standard deviation$^{71,72}$. We finally verify the coherence paradox (\ref{eqpd01}) and calculate the winning probability of the coherence game (\ref{eqpw}).

\section*{Supplemental Information}

Supplemental information can be found online

\section*{Acknowledgements}

We thank the discussion with Prof. Shao-Ming Fei. This work was supported by the National Natural Science Foundation of China (Nos. 62172341, 61772437), Sichuan Natural Science Foundation (No. 2023NSFSC0447), and Interdisciplinary Research of Southwest Jiaotong University China (No. 2682022KJ004).

\section*{Author Contributions}

Y.H. and M.X. conceived the study. Y.H. and M.X. designed the experiment. Y.H. and X.Z. conducted the experiments. Y.H., Y.X., and M.X. wrote the paper. All authors reviewed the paper.

\section*{Declaration of Interests}

The authors declare no competing interests.

\section*{Inclusion and Diversity}

We support inclusive, diverse, and equitable conduct of research.

\renewcommand \baselinestretch{0.5} \selectfont

\section*{Note S1. Generalized paradox for witnessing quantum coherence}

\subsection*{A. Generalized superposition}

Consider the general linear supposition of two quantum sources $|\psi_{01}\rangle$ and $|\psi_{10}\rangle$ as $|\psi_{00}(\theta)\rangle=\cos\theta|\psi_{01}\rangle+\sin\theta|\psi_{10}\rangle$ with $\theta\in (0,\pi/2)$. This implies a generalized coherence paradox
\begin{equation}
\left\{
\begin{aligned}
&\langle \sigma_Z^1 \sigma_Z^2\rangle_{|\psi_{01}\rangle} =-1,
\\
&\langle \sigma_Z^1 \sigma_Z^2\rangle_{|\psi_{10}\rangle} =-1,
\\
&\langle \sigma_X^1 \sigma_X^2\rangle_{|\psi_{01}\rangle} =0,
\\
&\langle \sigma_X^1 \sigma_X^2\rangle_{|\psi_{10}\rangle} =0,
\\
& \langle \sigma_X^1 \sigma_X^2\rangle_{|\psi_{00}(\theta)\rangle}=\sin(2\theta).
\end{aligned}
\right.
\label{eqpd02}
\end{equation}
Any LHV model cannot interpret this. In fact, assume $\lambda_{00}$ is a classical mixture of $\lambda_{01}$ and $\lambda_{10}$, i.e., $\lambda_{00}=p_1\lambda_{01}+p_2\lambda_{10}$ with a probability $\{p_1,p_2\}$. Both the equalities of $v^1v^2(xy=00)=0$, $v^1v^2(xy=01)=0$ imply $v^1v^2(xy=10)=0$ which implies  a sharp contradiction of $0=\sin(2\theta)$ for any $\theta\in (0,\pi/2)$.

Similarly, with Pauli $\sigma_Y$ and Pauli $\sigma_Z$ measurements, we have
\begin{equation}
\left\{
\begin{aligned}
&\langle \sigma_Z^1 \sigma_Z^2\rangle_{|\psi_{01}\rangle} =-1,
\\
&\langle \sigma_Z^1 \sigma_Z^2\rangle_{|\psi_{10}\rangle} =-1,
\\
&\langle  \sigma_Y^1 \sigma_Y^2\rangle_{|\psi_{01}\rangle}=0,
\\
&\langle  \sigma_Y^1 \sigma_Y^2\rangle_{|\psi_{10}\rangle}=0,
\\
&\langle  \sigma_Y^1\sigma_Y^2\rangle_{|\psi_{00}\rangle}=+1,
\end{aligned}
\right.
\label{eqpd03}
\end{equation}
for maximally entangled EPR state $|\psi_{00}\rangle=(|01\rangle+|10\rangle)/\sqrt{2}$. In the LHV model, we set $\lambda_{00}$ is classical mixture of $\lambda_{01}$ and $\lambda_{10}$. And then, we can infer the equality $v^1v^2(xy=00)=0$ from both equalities $v^1v^2(xy=01)=0$ and $v^1v^2(xy=10)=0$. This leads to a sharp contradiction of $0=+1$. Furthermore, both $\langle \sigma_Z^1 \sigma_Z^2 \rangle=-1$ and $\langle \sigma_Y^1 \sigma_Y^2 \rangle=0$ indicate the definite particle behaviors of the quantum source.

Consider the general linear supposition of two quantum sources $|\psi_{01}\rangle$ and $|\psi_{10}\rangle$ as $|\psi_{00}(\theta)\rangle$. The generalized coherence paradox is given by
\begin{equation}
\left\{
\begin{aligned}
&\langle \sigma_Z^1 \sigma_Z^2\rangle_{|\psi_{01}\rangle} =-1,
\\
&\langle \sigma_Z^1 \sigma_Z^2\rangle_{|\psi_{10}\rangle} =-1,
\\
&\langle  \sigma_Y^1 \sigma_Y^2\rangle_{|\psi_{01}\rangle}=0,
\\
&\langle  \sigma_Y^1 \sigma_Y^2\rangle_{|\psi_{10}\rangle}=0,
\\
&\langle  \sigma_Y^1 \sigma_Y^2\rangle_{|\psi_{00}(\theta)\rangle}=\sin(2\theta),
\end{aligned}
\right.
\label{eqpd04}
\end{equation}
which cannot be interpreted using any LHV model.

\subsection*{B. Multiple-slit interference}

Multiple-slit interference is a phenomenon in which light waves passing through multiple slits create an interference pattern. It is a variation of the famous Young double-slit experiment$^{33}$, where instead of two slits, there are multiple slits in a barrier. The interference pattern formed by multiple slits is more complex than that of two slits. The number of bright regions, known as fringes, increases with the number of slits. The behavior of light in multiple-slit interference can be explained using the concept of wave interference. Here, we provide a method to feature the kind of experiment using coherence paradox. Especially, we show the multiple coherence with $n$-partite Dicke state with one excitation which is given by$^{61}$
\begin{eqnarray}
|D_n^{(1)}\rangle=\frac{1}{\sqrt{n}}\sum_{\textsf{g}\in S_n} \textsf{g}(|0\rangle^{\otimes(n-1)}|1\rangle),
\end{eqnarray}
where the state is a superposition of all possible permutations of $n-1$ number of $|0\rangle$ and $|1\rangle$, and $S_n$ denotes the permutation group of $n$. Here, each slit will be presented by a state of $\textsf{g}(|0\rangle^{\otimes(n-1)}|1\rangle)$.

Consider the special cases of $|\phi_{000}\rangle=|D_3^{(1)}\rangle$. We define $|\phi_{001}\rangle= |001\rangle$, $|\phi_{010}\rangle= |010\rangle$, and $|\phi_{100}\rangle= |100\rangle$. The results of the Pauli measurements,
\begin{equation}
\left\{
\begin{split}
&\langle \sigma_Z^1\sigma_Z^2 \sigma_Z^3\rangle_{|\phi_{001}\rangle} =-1,
\\
&\langle \sigma_Z^1 \sigma_Z^2\sigma_Z^3\rangle_{|\phi_{010}\rangle} =-1,
\\
& \langle \sigma_Z^1 \sigma_Z^2 \sigma_Z^3\rangle_{|\phi_{100}\rangle}=-1,
\\
&\langle \sigma_X^1 \sigma_X^2 \sigma_Z^3\rangle_{|\phi_{001}\rangle} =0,
\\
&\langle \sigma_X^1 \sigma_X^2 \sigma_Z^3\rangle_{|\phi_{010}\rangle} =0,
\\
& \langle \sigma_X^1 \sigma_X^2 \sigma_Z^3\rangle_{|\phi_{100}\rangle}=0,
\\
& \langle \sigma_X^1 \sigma_X^2 \sigma_Z^3\rangle_{|\phi_{000}\rangle}= +\frac{2}{3}.
\end{split}
\right.
\label{eqpd05}
\end{equation}
In the LHV model$^3$, the variable $\lambda_{000}$ is a classical mixture of $\lambda_{001}$, $\lambda_{010}$, and $\lambda_{100}$ with any nontrivial probability distribution $\{p_1,p_2,p_3\}$. This follows the unique result of $v^1v^2v^3(xy=000)=0$ based on the equations $v^1v^2v^3(xy=001)=0$, $v^1v^2v^3(xy=010)=0$, and $v^1v^2v^3(xy=100)=0$. However, this does not match the last equality, i.e., $0=2/3$. So, this paradox shows both particle and wave features of input states. Another two paradoxes are given
\begin{equation}
\left\{
\begin{split}
&\langle \sigma_Z^1 \sigma_Z^2\sigma_Z^3\rangle_{|\phi_{001}\rangle} =-1,
\\
&\langle \sigma_Z^1 \sigma_Z^2 \sigma_Z^3\rangle_{|\phi_{010}\rangle} =-1,
\\
& \langle\sigma_Z^1 \sigma_Z^2 \sigma_Z^3\rangle_{|\phi_{100}\rangle}=-1,
\\
&\langle \sigma_X^1 \sigma_Z^2 \sigma_X^3\rangle_{|\phi_{001}\rangle} =0,
\\
&\langle \sigma_X^1\sigma_Z^2 \sigma_X^3\rangle_{|\phi_{010}\rangle} =0,
\\
& \langle \sigma_X^1 \sigma_Z^2 \sigma_X^3\rangle_{|\phi_{100}\rangle}=0,
\\
& \langle \sigma_X^1 \sigma_Z^2 \sigma_X^3\rangle_{|\phi_{000}\rangle}= +\frac{2}{3},
\end{split}
\right.
\label{eqpd06}
\end{equation}
and
\begin{equation}
\left\{
\begin{split}
&\langle \sigma_Z^1 \sigma_Z^2 \sigma_Z^3\rangle_{|\phi_{001}\rangle} =-1,
\\
&\langle \sigma_Z^1 \sigma_Z^2 \sigma_Z^3\rangle_{|\phi_{010}\rangle} =-1,
\\
& \langle \sigma_Z^1 \sigma_Z^2 \sigma_Z^3\rangle_{|\phi_{100}\rangle}=-1,
\\
&\langle \sigma_Z^1 \sigma_X^2 \sigma_X^3\rangle_{|\phi_{001}\rangle} =0,
\\
&\langle \sigma_Z^1 \sigma_X^2 \sigma_X^3\rangle_{|\phi_{010}\rangle} =0,
\\
& \langle \sigma_Z^1 \sigma_X^2 \sigma_X^3\rangle_{|\phi_{100}\rangle}=0,
\\
& \langle \sigma_Z^1 \sigma_X^2 \sigma_X^3\rangle_{|\phi_{000}\rangle}= +\frac{2}{3}.
\end{split}
\right.
\label{eqpd07}
\end{equation}

To sum up, for a Dicke state that contains one excitation in an $n$-partite system, it is always suitable to perform Pauli $\sigma_X$ measurement on $n-1$ particles of the system to construct a coherence paradox while the other is with Pauli $\sigma_Z$ as
\begin{equation}
\left\{
\begin{split}
&\textsf{g}(\langle  \sigma_Z^{\otimes{}n} \rangle)_{\textsf{g}(|0\rangle^{\otimes(n-1)}|1\rangle)} =-1,
\\
& \textsf{g}(\langle  \sigma_X^{\otimes{}n-1}\otimes \sigma_Z\rangle)_{\textsf{g}(|0\rangle^{\otimes(n-1)}|1\rangle)} =0,
\\
& \textsf{g}(\langle  \sigma_X^{\otimes{} n-1}\otimes \sigma_Z\rangle)_{|D_n^{(1)}\rangle}= +\frac{n-1}{n},
\end{split}
\right.
\label{eqpd05}
\end{equation}
where $\textsf{g}\in S_n$ denotes the permutation group of $n$. Within the LHV model,  the classical mixture of all variables implies the linear adding of all the values for $v^1v^2\cdots v^n$ associated with the corresponding input state $\textsf{g}(|0\rangle^{\otimes(n-1)}|1\rangle)$. This implies that $v^1v^2\cdots v^n=0$ for the input $0\cdots0$. This conflicts with the fact from the last equality, i.e., $v^1v^2\cdots v^n=(n-1)/n$.

\section*{Note S2. Experimental setup}

The schematic of the experimental setup is shown in the main text (Figure~2). The light source module contains a 405nm continuous-wave diode laser (Laser), a polarization beam splitter (PBS1), and two groups of wave plates. These devices operate at a wavelength of 405nm, and the laser emits successive pulses of this wavelength through the first wave plate group, 405nm polarization beam splitter, and the second wave plate group. The first wave plate group (QWP1+HWP1) is used with PBS1 to further adjust the phase and intensity of the laser, while the second wave plate group (QWP2+HWP2) regulates the intensity distribution in the Sagnac interference.

The Sagnac interference module consists of a dual-wavelength polarizing beam splitter (DPBS), a dual-wavelength half-wave plate (DHWP)  oriented at $45^\circ$, a quasi-phase-matched periodically poled KTiOPO$_4$ (PPKTP), and a polishing mirror set (M1, M2). The DPBS and DHWP operate at 405nm and 810nm. The PPKTP crystal is an artificial crystal that operates at a wavelength range of 400-4000nm, and for this module, it operates at 405nm to 810nm at a temperature range of $15^\circ$C-$80^\circ$C. The laser generates linearly polarized pump light, which then passes through the Sagnac interference module to create a polarization superposition state represented by the equation $\cos\theta |H\rangle+e^{i\varphi}\sin\theta|V\rangle$. The parameters $\theta$ can be set by the rotation of the HWP, and the phase angle $\varphi$ can be cleared by adjusting the optical axis direction of the QWP. Once passing through DM and DPBS, the horizontal pump light is focused on the PPKTP crystal, which results in the production of a clockwise down-converted 810 nm photon pair $|HV\rangle$. This pair of photons then becomes $|VH\rangle$ after passing through DPBS. On the other hand, the vertical pump light is rotated to a horizontal orientation by DHWP and focused on the PPKTP crystal, resulting in a counter-clockwise photon pair $|HV\rangle$. After passing through DPBS once again, the horizontal photons are transmitted while the vertical photons are reflected. Then, the photons from path-I and -II are further coupled into a single-mode fiber, respectively, forming a polarization-entangled Bell-like state.

The projection measurement module comprises two identical optical paths, each with the same components. These components include an 810nm quarter-wave plate, an 810nm half-wave plate, an 810nm polarization beam splitter, and an interference filter. The Sagnac interference is split into two optical paths. Upon being separated by DM, some of the 405nm pump light in path-I is reflected along its original route until it reaches PBS1, where it is directed out of the optical path. After passing through each 810nm PBS, there is an 800nm filter, which allows for high light transmission above 800nm while blocking the remaining light with high reflectivity.

\section*{Note S3. State tomography and experimental data}

\begin{figure*}[htbp]
\centering
\subfigure[$\theta=0$]{
\begin{minipage}[t]{0.45\linewidth}
\centering
\includegraphics[width=2.8in]{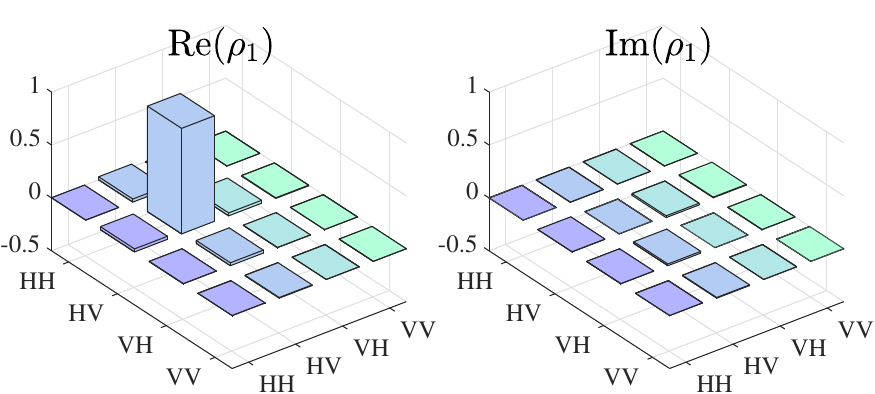}
\end{minipage}%
}%
\subfigure[$\theta=\frac{\pi}{12}$]{
\begin{minipage}[t]{0.45\linewidth}
\centering
\includegraphics[width=2.8in]{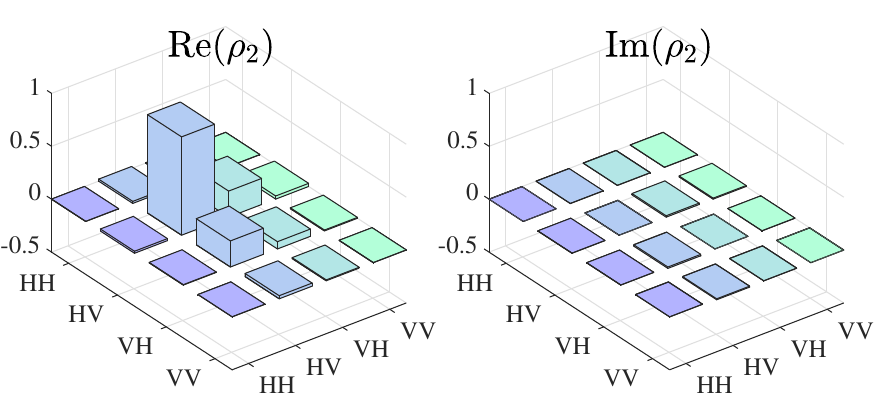}
\end{minipage}%
}%
\quad
\subfigure[$\theta=\frac{\pi}{8}$]{
\begin{minipage}[t]{0.45\linewidth}
\centering
\includegraphics[width=2.8in]{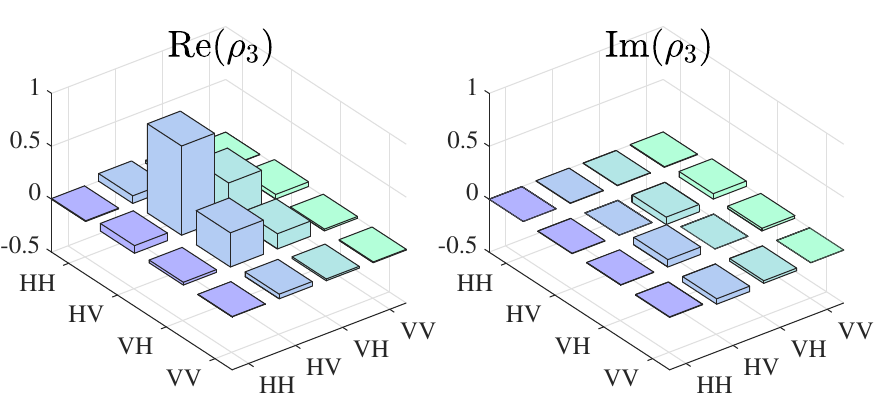}
\end{minipage}
}%
\subfigure[$\theta=\frac{\pi}{6}$]{
\begin{minipage}[t]{0.45\linewidth}
\centering
\includegraphics[width=2.8in]{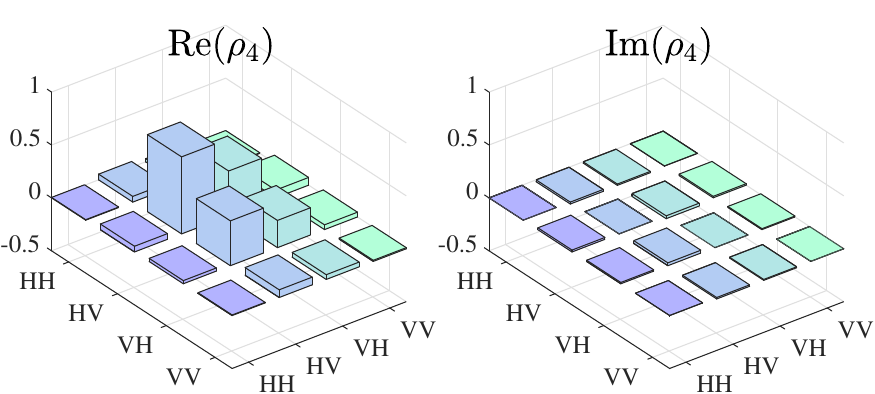}
\end{minipage}
}%
\quad
\subfigure[$\theta=\frac{\pi}{4}$]{
\begin{minipage}[t]{0.45\linewidth}
\centering
\includegraphics[width=2.8in]{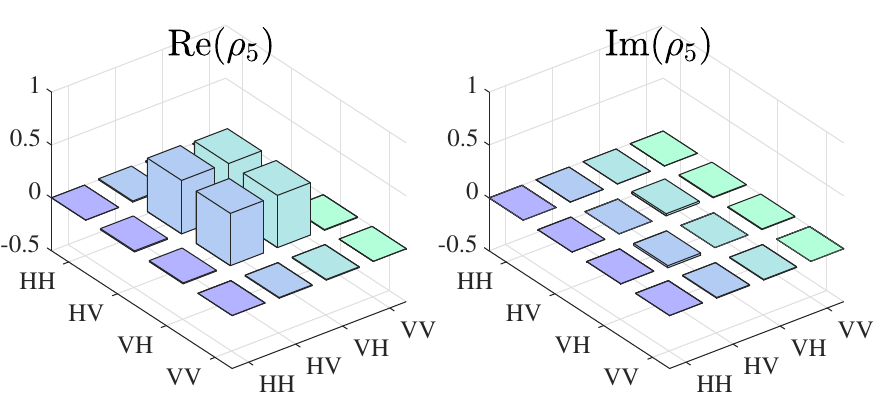}
\end{minipage}
}%
\subfigure[$\theta=\frac{\pi}{2}$]{
\begin{minipage}[t]{0.45\linewidth}
\centering
\includegraphics[width=2.8in]{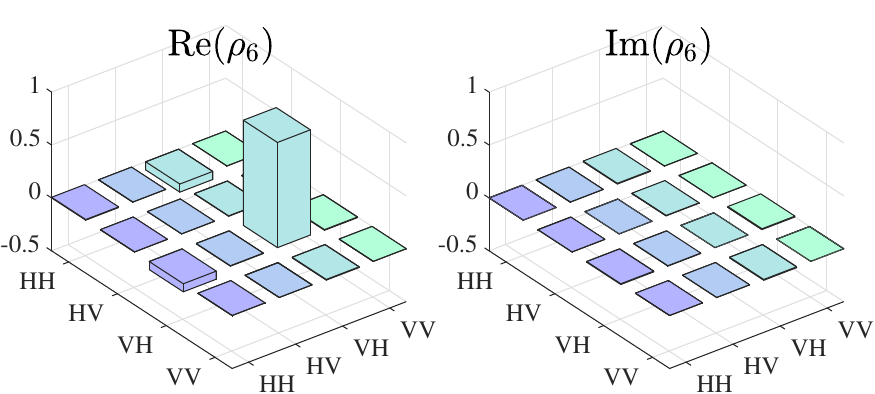}
\end{minipage}
}%
\centering
\caption{The tomographic results for the six states. (a) $\rho_1=|\psi_{01}\rangle \langle\psi_{01}|$. (b) $\rho_2=|\psi_{00}(\pi/12)\rangle \langle\psi_{00}(\pi/12)|$. (c) $\rho_3=|\psi_{00}(\pi/8)\rangle \langle\psi_{00}(\pi/8)|$. (d) $\rho_4=|\psi_{00}(\pi/6)\rangle \langle\psi_{00}(\pi/6)|$. (e) $\rho_5=|\psi_{00}(\pi/4)\rangle \langle\psi_{00}(\pi/4)|$. (f) $\rho_6=|\psi_{10}\rangle \langle\psi_{10}|$. The ``Re'' represents the real part of the density matrices and the ``Im'' for the imaginary part.}
\label{FIGSI}
\end{figure*}

With the standard state tomography methods for linear optics$^{30,60}$, we get six experimental states  $|\psi_{01}\rangle$, $|\psi_{00}({\pi}/{12})\rangle$, $|\psi_{00}({\pi}/{8})\rangle$, $|\psi_{00}({\pi}/{6})\rangle$, $|\psi_{00}({\pi}/{4})\rangle$, and $|\psi_{10}\rangle$, respectively (Figure~\ref{FIGSI}). We can see that all the states have high fidelity with small imaginary parts, which ensures the large violation of the coherence paradox.

\begin{figure}[h!]
\begin{center}
\includegraphics[width=0.5\linewidth]{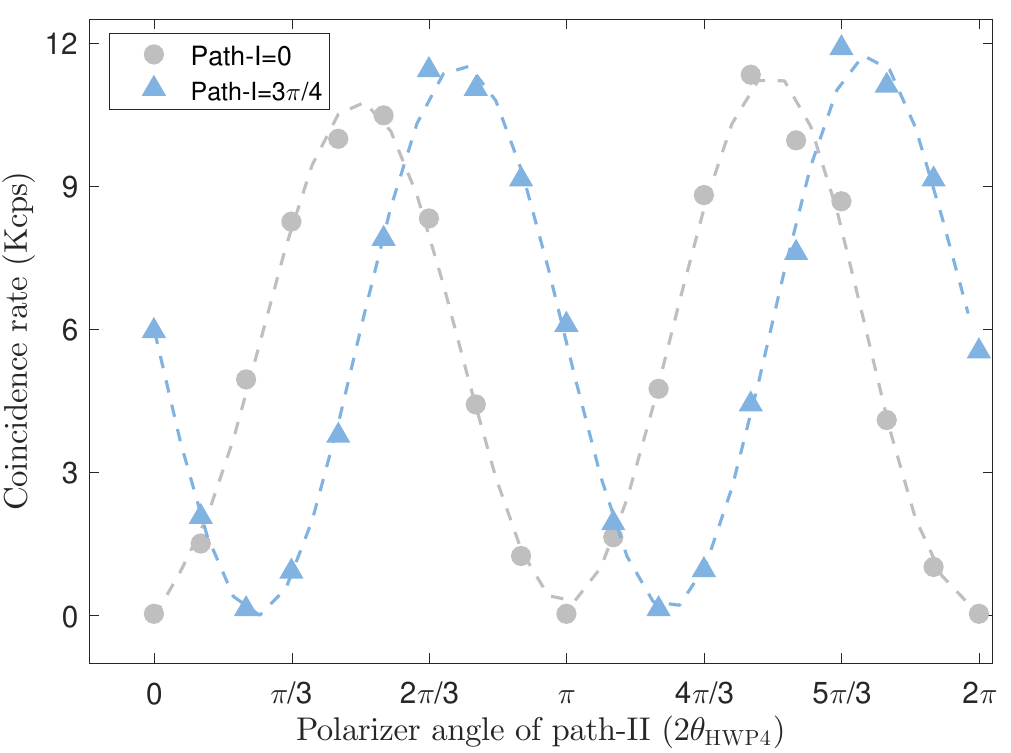}
\caption{Experimentally coincidence rate of path-I (the half HWP3-angle) vs. polarizer angle of path-II (the half HWP4-angle). The blue part represents path-I fixed at angle 0, while the red represents path-I fixed at angle $3\pi/4$.  Here, HWP plus PBS in front of detectors act as a polarizer box such that $\theta_{\rm{pol}}=2\theta_{\rm{HWP4}}$.}
\label{FigureIF}
\end{center}
\end{figure}

We have plotted the behavior of the experimentally measured coincidence rate when the polarizer angle of path-I (or half of the HWP3 angle) is fixed at angles 0 and $3\pi/4$ for the different values of the polarizer angles of path-II that agrees with the theoretical quantum prediction (Figure~\ref{FigureIF}). We obtain $V_{HV}=\%(0.9966\pm 0.0008)$ and $V_{DA}=\%( 0.9802\pm 0.0020)$.  The measured visibility shows a very strong and reliable violation from the classical physics prediction ($V\leq \%71$).

Table \ref{paradox2} presents experimental results on the coherence paradox under the post-selection of 8 sets of coincidences ($\theta={\pi}/{12}, {\pi}/{8}, {\pi}/{6}, {\pi}/{4}$), as described by the generalized GHZ-type paradox (\ref{eqpd02}) with Pauli matrices $\sigma_X$ and $\sigma_Z$. Despite the existence of the noise, the experimental values confirm that the generalized entangled two-photon EPR state violates the present GHZ-type paradox. Especially, experimental values of $\langle \sigma_X^1 \sigma_X^2\rangle_{|\psi_{00}\rangle}$ are almost more than 10 times larger than $\langle \sigma_X^1 \sigma_X^2\rangle_{|\psi_{01}\rangle}$ or $\langle \sigma_X^1 \sigma_X^2\rangle_{|\psi_{10}\rangle}$. This observation cannot be explained by any local hidden-variable (LHV) model following the proof procedure of the paradox (\ref{eqpd02}).

\begin{table}[h!]
    \centering
        \caption{Experimental values of correlations concerning the generalized coherence paradox with Pauli $\sigma_X$ and Pauli $\sigma_Z$ measurements.}
    \begin{tabular}{l|l|c|c}
   \hline
       Correlators  & Experimental value & Theoretical value & Validity   \\ \hline
       $\langle \sigma_Z^1 \sigma_Z^2\rangle_{|\psi_{01}\rangle}$  & $ -0.9967 \pm 0.0005$ & -1 & \checkmark
\\
        $\langle \sigma_Z ^1\sigma_Z^2\rangle_{|\psi_{10}\rangle}$  & $ -0.9912 \pm 0.0007$ & -1 & \checkmark
\\
         $\langle \sigma_X^1 \sigma_X^2\rangle_{|\psi_{01}\rangle}$  & $ 0.0625 \pm 0.0084$ & 0 & \checkmark
\\
         $\langle \sigma_X^1 \sigma_X^2\rangle_{|\psi_{10}\rangle}$  & $ 0.0317 \pm 0.0080$ & 0 & \checkmark
\\
         $\langle \sigma_X^1 \sigma_X^2\rangle_{|\psi_{00}({\pi}/{12})\rangle}$  & $ 0.4944 \pm 0.0081$ & +0.5000 & \checkmark
\\
         $\langle \sigma_X^1 \sigma_X^2\rangle_{|\psi_{00}({\pi}/{8})\rangle}$  & $  0.7076 \pm 0.0057$ & +0.7071 & \checkmark
\\
         $\langle \sigma_X^1 \sigma_X^2\rangle_{|\psi_{00}({\pi}/{6})\rangle}$  &$0.8685 \pm 0.0039$ & +0.8660 & \checkmark
\\
         $\langle \sigma_X^1 \sigma_X^2\rangle_{|\psi_{00}({\pi}/{4})\rangle}$  & $0.9949 \pm 0.0006$ & +1 & \checkmark
\\  \hline
    \end{tabular}
    \label{paradox2}
\end{table}

Using Pauli matrix $\sigma_Y$ and $\sigma_Z$ as local measurements described in the paradox (\ref{eqpd04}), we conducted tests on 8 sets of coincidences ($\theta={\pi}/{12}, {\pi}/{8}, {\pi}/{6}, {\pi}/{4}$) under a post-selection. It verified that the generalized entangled two-photon EPR state violates the present GHZ-type paradox. Moreover, the observation of $\langle \sigma_X^1 \sigma_X^2\rangle_{|\psi_{00}\rangle}$ is more than 10 times larger than $\langle \sigma_X^1 \sigma_X^2\rangle_{|\psi_{01}\rangle}$ or $\langle \sigma_X^1 \sigma_X^2\rangle_{|\psi_{10}\rangle}$, which cannot be explained in any LHV model, that is, the classical mixture of the measured results of independent states cannot represent the measured results of the superposition states.
\begin{table}[h!]
    \centering
        \caption{Experimental values of correlations concerning the generalized coherence paradox with Pauli $\sigma_Y$ and Pauli $\sigma_Z$ measurements.}
    \begin{tabular}{l|c|c|c}
    \hline
      Correlators & Experimental value & Theoretical value & Validity   \\ \hline
       $\langle \sigma_Z^1 \sigma_Z^2\rangle_{|\psi_{01}\rangle}$  & $ -0.9967 \pm 0.0005$ & -1 & \checkmark
\\
        $\langle \sigma_Z^1 \sigma_Z^2\rangle_{|\psi_{10}\rangle}$  & $ -0.9912 \pm 0.0007$ & -1 & \checkmark
\\
         $\langle \sigma_Y^1 \sigma_Y^2\rangle_{|\psi_{01}\rangle}$  & $ 0.0625 \pm 0.0084$ & 0 & \checkmark
\\
         $\langle \sigma_Y^1 \sigma_Y^2\rangle_{|\psi_{10}\rangle}$  & $0.0317 \pm 0.0080$ & 0 & \checkmark
\\
         $\langle \sigma_Y^1 \sigma_Y^2\rangle_{|\psi_{00}({\pi}/{12})\rangle}$  & $ 0.4937 \pm 0.0100$ & +0.5000 & \checkmark
\\
         $\langle \sigma_Y^1 \sigma_Y^2\rangle_{|\psi_{00}({\pi}/{8})\rangle}$  & $0.7203 \pm 0.0083$ & +0.7071 & \checkmark
\\
         $\langle \sigma_Y^1 \sigma_Y^2\rangle_{|\psi_{00}({\pi}/{6})\rangle}$  &$0.8566 \pm 0.0052$ & +0.8660 & \checkmark
\\
         $\langle \sigma_Y^1 \sigma_Y^2\rangle_{|\psi_{00}({\pi}/{4})\rangle}$  & $ 0.9885 \pm 0.0011$ & +1 & \checkmark
\\  \hline
    \end{tabular}
    \label{paradox3}
\end{table}

By utilizing Eq.~(14) and photon correlations (16) shown in the main text, we assessed the experimental winning probability of the coherence game. The theoretical winning probability can achieve a maximum violation of $P_{win}^{Q}=5/8$, surpassing $P_{win}^{C}=1/2$, when two players measure the states with either Pauli measurement  $\sigma_X$ or $\sigma_Z$. The experimental results are presented in Table \ref{pwin1} and \ref{pwin2} accordingly. Notably, we observed a maximal violation with the maximally entangled EPR state, revealing a quantum superposition that surpasses any classical correlations between the two players.

\begin{table}[h!]
    \centering
        \caption{Experimental estimations of the winning probability of quantum XOR game under Pauli $\sigma_X$ measurement per party.}
{\small   \begin{tabular}{l|c|c|c|c|c|c}
    \hline
       \multirow{2}*{State}  & Experimental  & Theoretical & Experimental & Theoretical & Theoretical & \multirow{2}*{Validity}
\\
      &value  of $\langle \sigma_X^1 \sigma_X^2\rangle$ &value of $\langle \sigma_X^1 \sigma_X^2\rangle$ &value of $P_{win}^Q$ &value of $P_{win}^Q$ &value of $P_{win}^C$ &
\\ \hline
       $|\psi_{01}\rangle$  & $  0.0625 \pm 0.0084$ & 0 & - & - & - & \checkmark
\\
        $|\psi_{10}\rangle$  & $  0.0317 \pm 0.0080$ & 0 & - & - & - & \checkmark
\\
         $|\psi_{00}(\pi/12)\rangle$  & $ 0.4944 \pm 0.0081$ & 0.5000 & $0.5500 \pm 0.0031$ & 0.5625 & 0.5000 & \checkmark
\\
         $|\psi_{00}(\pi/8)\rangle$  & $  0.7076 \pm 0.0057$ & 0.7071 & $0.5767 \pm 0.0028$ & 0.5884 & 0.5000 & \checkmark
\\
         $|\psi_{00}(\pi/6)\rangle$  & $ 0.8685 \pm 0.0039$ & 0.8660 & $0.5968 \pm 0.0025$ & 0.6083 & 0.5000 & \checkmark
\\
         $|\psi_{00}(\pi/4)\rangle$  & $ 0.9949 \pm 0.0006$ & 1 & $0.6126 \pm 0.0021$ & 0.6250 & 0.5000 & \checkmark
\\  \hline
    \end{tabular}
    }
    \label{pwin1}
\end{table}

\begin{table}[]
    \centering
\caption{Experimental estimations of the winning probability of quantum XOR game under the Pauli $\sigma_Z$ measurement per party.}
 {\small \begin{tabular}{l|c|c|c|c|c|c}
    \hline
       \multirow{2}*{State}  & Experimental & Theoretical & Experimental & Theoretical  & Theoretical & \multirow{2}*{Validity}
\\
      & value of $\langle \sigma_Z^1 \sigma_Z^2\rangle$ & value of $\langle \sigma_Z^1 \sigma_Z^2\rangle$ & value of $P_{win}^Q$ & value of $P_{win}^Q$ & value of $P_{win}^C$ &
\\ \hline
       $|\psi_{01}\rangle$  & $ -0.9967 \pm 0.0005$ & -1 & - & - & - & \checkmark
\\
        $|\psi_{10}\rangle$  & $ -0.9912 \pm 0.0007$ & -1 & - & - & - & \checkmark
\\
         $|\psi_{00}({\pi}/{12})\rangle$  & $ -0.9958 \pm 0.0007$ & -1 & $0.6240 \pm 0.0002$ & 0.6250 & 0.5000 & \checkmark
\\
         $|\psi_{00}({\pi}/{8})\rangle$  & $ -0.9806 \pm 0.0013$ & -1 & $0.6259 \pm 0.0003$ & 0.6250 & 0.5000 & \checkmark
\\
         $|\psi_{00}({\pi}/{6})\rangle$  & $ -0.9777 \pm 0.0014$ & -1 & $0.6263 \pm 0.0003$ & 0.6250 & 0.5000 & \checkmark
\\
         $|\psi_{00}({\pi}/{4})\rangle$  & $ -0.9973 \pm 0.0004$ & -1 & $0.6238 \pm 0.0002$ & 0.6250 & 0.5000 & \checkmark
\\  \hline
    \end{tabular}
    }
    \label{pwin2}
\end{table}

\end{document}